\documentclass[aps,prd,reprint,groupedaddress,amsmath,amssymb,amsfonts]{revtex4-2}

\usepackage[
  colorlinks=true,
  linkcolor=blue,
  citecolor=blue,
  urlcolor=black
]{hyperref}

\usepackage{subfigure}
\usepackage{graphicx}
\usepackage{hyperref}
\usepackage{physics}
\usepackage{natbib} 
\usepackage{xcolor}
\usepackage{amsfonts}

\begin{document}

\title{ Classical emergence of the quantum-backreacted BTZ black hole from exponential electrodynamics}

\author{ Julio A. M\'endez-Zavaleta}
\email{ julmendez@uv.mx}
\affiliation{Facultad de F\'isica, Universidad Veracruzana,  Paseo No. 112, Desarrollo Habitacional Nuevo Xalapa, C.P. 91097, Xalapa-Enr\'iquez, M\'exico.}

\author{ Efra\'in Rojas}
\email{efrojas@uv.mx}
\affiliation{Facultad de F\'isica, Universidad Veracruzana,  Paseo No. 112, Desarrollo Habitacional Nuevo Xalapa, C.P. 91097, Xalapa-Enr\'iquez, M\'exico.}

\author{  Jos\'e Joaquín Su\'arez-Garibay}
\email{zs24019372@estudiantes.uv.mx }
\affiliation{Facultad de F\'isica, Universidad Veracruzana,  Paseo No. 112, Desarrollo Habitacional Nuevo Xalapa, C.P. 91097, Xalapa-Enr\'iquez, M\'exico.}

\date{ \today}

\begin{abstract}  
   In this work, we revisit a recently reported generalization of the Ba\~nados--Teitelboim--Zanelli black hole arising in New Massive Gravity sourced by the quantum fluctuations of scalar matter, now examined through the lens of a purely classical framework. We show that the same geometry, distinguished by its logarithmic asymptotic structure, emerges as the unique static solution of Einstein gravity coupled to an exponential nonlinear electrodynamics. We trace the origin of this correspondence and prove that this geometry belongs to a unique class of metrics constituting the intersection of the moduli spaces of the static and circularly symmetric sectors of the two theories, thereby revealing a dynamical equivalence between them. An explicit mapping is established between the global charges of the nonlinearly charged black holes and the parameters governing the quantum backreaction in New Massive Gravity, allowing for a natural reinterpretation of the quantum imprints in terms of classical charges. A detailed analysis of the horizon structure of these spacetimes is presented. In addition, the full thermodynamics of the more general configurations is constructed using the Iyer--Wald formalism, from which we derive the first law and the associated Smarr relation. Altogether, our results provide a classical realization of a semiclassical spacetime and point toward a broader correspondence between higher-curvature corrections in quantum gravity and nonlinear effects in self-gravitating electrodynamics in three dimensions.
\end{abstract}

\maketitle
 

\section{Introduction}

In the absence of a fully developed theory of Quantum Gravity (QG), various strategies have been pursued to characterize the ultraviolet behavior of the gravitational field. For instance, quantum-backreacted spacetimes emerge when the effective energy-momentum tensor of a quantum field theory (QFT) is coupled to a classical set of field equations for the gravitational field. This semiclassical approach, though approximate, has proven valuable in capturing leading-order quantum effects in gravitational backgrounds, particularly around black holes, including perturbative one-loop corrections to classical geometries and their phenomenological implications \cite{Battista:2023iyu}.

From the gravitational dynamics side, quantum corrections are usually modeled by introducing higher-curvature terms as perturbative corrections to General Relativity (GR). New Massive Gravity (NMG) stands out as a compelling framework defined in three spacetime dimensions that implements this idea in a controlled and tractable manner. Introduced by Bergshoeff, Hohm, and Townsend in \cite{Bergshoeff:2009aq}, NMG modifies standard gravity by including specific combinations of quadratic curvature invariants that give rise to a parity-preserving theory of a massive graviton. Around maximally symmetric backgrounds, the theory remains ghost-free at the linearized level. These features, together with its rich solution space ---including exact black holes with weakened asymptotic behavior, see for instance \cite{Oliva:2009ip,Clement:2009gq,Ayon-Beato:2009rgu,Ayon-Beato:2015jga,Bravo-Gaete:2020ftn}---make NMG a fertile ground for exploring the interplay between higher-curvature gravity and quantum matter effects.

The action of this theory consists of the standard Einstein-Hilbert term with a cosmological constant $\lambda$, supplemented by a specific higher-curvature correction that ensures the theory propagates the degrees of freedom of a single massive spin-2 mode. Namely,
\begin{align}
S[g_{\mu\nu}] = \frac{1}{16\pi G} \int d^3x \sqrt{-g} \left( R - 2\lambda + \frac{1}{m^2} K \right),
\end{align}
where $ K $ is the curvature-squared correction term defined as  
\begin{equation} \label{eq:Kterm}
K = R_{\mu\nu} R^{\mu\nu} - \frac{3}{8} R^2,
\end{equation}
constructed from the Ricci tensor $R_{\mu\nu}$ and the scalar curvature $R$, and where $m$ is a mass parameter that controls the excitation scale of the massive graviton.

Unlike Topologically Massive Gravity (TMG) \cite{Deser:1982vy}, a former three-dimensional candidate for massive gravity based on the inclusion of a gravitational Chern-Simons term, NMG achieves the emergence of mass without sacrificing parity invariance. In TMG, the parity-odd Chern-Simons contribution introduces third-order derivatives in the field equations and leads to a chiral spectrum, where only one of the graviton helicities propagates depending on the sign of the coupling. This feature gives rise to a tension between bulk unitarity and the positivity of the boundary central charge. In this sense, NMG offers a more symmetric and appealing higher-curvature alternative.

In this context, a recent result presented in \cite{Chernicoff:2024dll,Chernicoff:2024lkj} explores the interplay between higher-curvature corrections to GR and quantum matter sources. In particular, a novel static geometry is obtained as the quantum backreaction induced by a large number of conformally coupled scalar fields. 
This is achieved by sourcing the quadratic curvature corrections of NMG with the renormalized expectation value of the energy-momentum tensor $\langle T_{\mu\nu} \rangle$ in a semiclassical setup \cite{Parker1968,Parker1969_QFPEI,Parker1971_QFPEII,DeWitt1975_QFTCS,Christensen1976_PointSplitting,Christensen1978_GeodesicPointSeparation}\footnote{Classical gravity sourced by $\langle T_{\mu\nu}\rangle$ has a long history, beginning with Parker’s analysis of particle creation, DeWitt’s covariant renormalization program, and Christensen’s point-splitting techniques. These developments form the modern basis of semiclassical gravity.}. The resulting gravitational system is governed by the following field equations
\begin{equation} 
R_{\mu\nu}-\frac{1}{2}Rg_{\mu\nu}+\lambda g_{\mu\nu}-\frac{1}{2m^2}K_{\mu\nu}=8\pi G N\langle T_{\mu\nu}\rangle, \label{eq:SemiclassicalNMGEqs}
\end{equation}
where
\begin{align}
    K_{\mu\nu}={}&2\nabla^2R_{\mu\nu}-\frac{1}{2}(\nabla_\mu \nabla_ \nu+g_{\mu\nu}\nabla^2)R-8R_{\mu\rho}R^{\rho}_{\ \ \nu}\\&+\frac{9}{2}RR_{\mu\nu}+\frac{1}{8}g_{\mu\nu}\left(24R^{\alpha \beta }R_{\alpha \beta}-13R^2\right),\nonumber
\end{align}
comes from the variation of the quadratic term \eqref{eq:Kterm}.  
The underlying matter constituent of $\langle T_{\mu\nu}\rangle$ is a conformally coupled scalar field propagating over the BTZ background geometry, whose dynamics is governed by the generalized Klein--Gordon equation, invariant under Weyl rescalings,
\begin{equation}
\left(\Box - \frac{1}{8} R \right)\psi = 0,
\end{equation}
and with classical energy-momentum tensor given by
\begin{align} \label{eq:Tmn_psi}
T_{\mu\nu} 
={}& \frac{3}{4}\,\nabla_\mu \psi \, \nabla_\nu \psi
- \frac{1}{4}\,\psi \, \nabla_\mu \nabla_\nu \psi
\\
&- \frac{1}{4}\, g_{\mu\nu}
\left[
\nabla_\alpha \psi \, \nabla^\alpha \psi
- \frac{\Lambda}{4}\, \psi^2
\right].\nonumber
\end{align}
In the setting described by Eq.~\eqref{eq:SemiclassicalNMGEqs}, the total matter sector is taken to consist of $N$ identical copies of this conformal field. The semiclassical source is therefore given by the renormalized expectation value of the energy-momentum tensor \eqref{eq:Tmn_psi}, whose quantum fluctuations collectively generate the backreaction on the geometry. Concretely,
\begin{equation} \label{eq:q-Tem}
\langle T^\mu_{\ \ \nu}\rangle=\frac{\hbar F(M)}{r^3}
\left(\delta_t^\mu \delta_\nu^t+\delta_r^\mu \delta_\nu^r-2 \delta_\phi^\mu \delta_\nu^\phi\right).
\end{equation}
This expression follows from applying a point-splitting regularization procedure to the two-point function of the quantum field, evaluated via the method of images on the Ba\~nados--Teitelboim--Zanelli (BTZ) background. Here $G$ denotes the three-dimensional Newton's gravitational constant. In this construction, a semiclassical large--$N$ regime is assumed, so that the quantum fluctuations of the matter fields provide the dominant contribution to the gravitational backreaction, while graviton fluctuations remain subleading.

This approach yields a finite, state-dependent expression that captures the leading corrections, following the original analysis in \cite{Martinez:1996uv}. The function $ F(M) $ appearing in \eqref{eq:q-Tem} is a smooth and bounded function of the BTZ mass $M$, and becomes exponentially suppressed for large $M$. Nevertheless, its backreaction leads to a nontrivial deformation of the classical geometry and plays a central role in modifying the asymptotic behavior of the solution.

The resulting semiclassical system admits a novel static black hole solution that generalizes the BTZ geometry through a logarithmic deformation. Derived in \cite{Chernicoff:2024dll}, the metric takes the form
\begin{subequations}\label{eq:solution}
\begin{align}
ds^2 &= -f(r)\,dt^2 + \frac{dr^2}{f(r)} + r^2\,d\phi^2, \label{eq:AnsatzMetric} \\
f(r) &= \frac{r^2}{l^2} - M + \frac{2N\ell_p F(M)}{M l^2} \left(r \ln{\left(\frac{r}{l}\right)} - r \right), \label{eq:fFunctionByMGT}
\end{align}
\end{subequations}
where $\ell_p = 8\pi G\hbar$ is the three-dimensional Planck length, and $l$ is the Anti-de Sitter (AdS) radius. The deformation is governed by the smooth function $F(M)$ and gives rise to a geometry that asymptotically deviates from the standard AdS$_3$ behavior. The modified boundary conditions encoded in \eqref{eq:AnsatzMetric}–\eqref{eq:fFunctionByMGT} lead to an enhanced asymptotic symmetry algebra featuring logarithmic supertranslations \cite{Chernicoff:2024dll}, and have been further explored in connection with black hole microstate constructions and holographic $c$-theorems in higher-curvature gravity \cite{Chernicoff:2024lkj}.

In this work, we show that the very same geometry can be realized as the unique solution of a purely classical system. Specifically, we demonstrate that the metric \eqref{eq:solution} emerges as the exact and unique static solution to Einstein gravity coupled to a family of nonlinear electrodynamics with exponential structure. This realization of the solution not only provides a new classical interpretation of the backreacted geometry and its parameters, but also uncovers a dynamical equivalence between the two seemingly distinct theoretical frameworks.
\\

\emph{Organization of this work.}  
In Sec.~\ref{sec:ExactSolutions} we introduce the Einstein–Nonlinear Electrodynamics (ENLED) model and discuss its general dynamical structure.  
In Sec.~\ref{sec:static_sols} we derive a new family of static solutions that includes the quantum-backreacted BTZ geometry within a purely classical framework.  
Sec.~\ref{sec:duality} is devoted to showing that, under the assumption of staticity and circular symmetry, the full solution space of the ENLED system coincides with that of semiclassical NMG, thereby establishing a dynamical equivalence between the two settings.  
The horizon structure and causal properties of the resulting geometries are analyzed in Sec.~\ref{sec:horizons}
In Sec.~\ref{sec:Thermo} we study the thermodynamics of the general solution: Sec.~\ref{sec:charges} focuses on the construction of the conserved charges and the verification of the extended first law and Smarr relation, while Sec.~\ref{sec:stability} addresses local thermodynamic stability.  
In Sec.~\ref{sec:classical_lim} we discuss how the nonlinearly charged family admits a natural interpretation as the classical limit of the quantum-backreacted BTZ configuration.  
Concluding remarks are presented in Sec.~\ref{sec:conclusions}.  
Appendix~\ref{appx:Iyerwaldformalism} provides details on the Iyer–Wald formalism employed in the thermodynamic analysis, while Appendix~\ref{appx:Plebanski} presents the equivalent electrodynamic description in the Plebanski $\mathcal H(\mathcal{P})$ formulation.

\section{Exact solutions in three-dimensional nonlinear electrodynamics \label{sec:ExactSolutions} }

Although pure AdS$_3$ gravity possesses no local propagating degrees of freedom, quantum fields on the BTZ background \cite{BTZ1992} generate a non-vanishing renormalized stress tensor. Early work by Martínez and Zanelli \cite{Martinez:1996uv} showed that this backreaction induces logarithmic corrections to the BTZ mass function and to the horizon position, a feature later revisited in analyses of quantum-corrected thermodynamics and rotating backgrounds \cite{Vagenas2001,KothawalaShankaSriram2008}. More recently, the results of \cite{Chernicoff:2024dll,Chernicoff:2024lkj} placed this mechanism within the framework of New Massive Gravity: sourcing the NMG equations with $\langle T_{\mu\nu}\rangle$ for conformally coupled scalars yields a static black hole with weakened AdS$_3$ asymptotics and a characteristic logarithmic sector. Taken together with the classical constructions discussed below, these observations suggest that nonlinear electromagnetic fields may offer a purely classical route to geometries that share structural features with their semiclassical counterparts.

Nonlinear electrodynamics (NLED) provides a parallel line of developments in three-dimensional gravity, supplying a remarkably rich catalogue of black-hole geometries that differ, in essential ways, from their Einstein–Maxwell analogues. Since the pioneering Born–Infeld proposal \cite{Born1934,BornInfeld1934,BornInfeld1935}, self-gravitating NLED has been explored in depth, with renewed momentum after the work of Ayón–Beato and García, who constructed the first fully regular black holes supported by a nonlinear electromagnetic field \cite{AyonBeatoGarcia1998,AyonBeatoGarcia1999}. This initiated a broad program aimed at understanding the geometry, thermodynamics, and photon propagation of NLED solutions, including general regular models \cite{Bronnikov2001} and Bardeen-type constructions \cite{Hayward2006}. In three dimensions, the role of NLED becomes particularly distinctive: because Maxwell theory does not produce a Coulomb-type potential, genuinely charged BTZ-like geometries arise only in the nonlinear regime. This was demonstrated by Cataldo and García \cite{Cataldo:2000ns}, who constructed $2+1$-dimensional charged black holes that reduce to BTZ in the linear limit yet remain regular throughout their core.

Let us consider GR minimally coupled to nonlinear electrodynamics (NLED). In contrast to the setup explored in \cite{Chernicoff:2024dll}, the additional degrees of freedom here arise from an external vector gauge field $A_\mu$, which in $D=3$ dimensions propagates a single physical degree of freedom per spacetime point. Explicitly, we consider the action
\begin{align} 
    S[g_{\mu\nu},A_\mu] = \frac{1}{2\kappa} \int d^3x\, \sqrt{-g} \left[ R - 2\lambda + 2\kappa \mathcal{L}(\mathcal{F}) \right],
    \label{eq:ActionEHNLED}
\end{align}
where $\lambda$ is the cosmological constant and $\kappa=8\pi G$ denotes Einstein’s gravitational constant in three dimensions. The function $\mathcal{L}$ is the Lagrangian density of the NLED sector, assumed to be an arbitrary, smooth, and differentiable function of the Maxwell invariant
\begin{equation}
    \mathcal{F} = \frac{1}{4} F_{\mu\nu} F^{\mu\nu}.
    \label{eq:InvariantF}
\end{equation}

Accordingly, $F_{\mu\nu}$ is the field-strength tensor of the gauge field, defined as the antisymmetrized derivative $F_{\mu\nu} := \partial_\mu A_\nu - \partial_\nu A_\mu$. This structure preserves the local $U(1)$ gauge invariance characteristic of electrodynamics. Alternatively, one may start from the equivalent Plebanski's formulation of the theory defined by \eqref{eq:InvariantF}, which provides a description in terms of a Hamiltonian-like structural function. This approach is briefly addressed in Appx.~\ref{appx:Plebanski}.

Varying the action \eqref{eq:ActionEHNLED} with respect to the metric yields the Einstein field equations coupled to a generic nonlinear electrodynamics
\begin{equation}
    R_{\mu\nu} - \frac{1}{2} R g_{\mu\nu} + \lambda g_{\mu\nu} = \kappa T_{\mu\nu},
    \label{eq:EFENLED}
\end{equation}
where the energy-momentum tensor takes the form
\begin{equation}
    T_{\mu\nu} = \mathcal{L}\, g_{\mu\nu}  - \mathcal{L}_{\mathcal{F}} F_{\mu\alpha} F_{\nu}^{\ \alpha}.
    \label{EnergyMomentumTensorNLED}
\end{equation}
Here, we have introduced the shorthand notation $\mathcal{L}_{\mathcal{F}} := \partial \mathcal{L} / \partial \mathcal{F}$.

Similarly, variation of Eq.~\ref{eq:ActionEHNLED} with respect to the electromagnetic potential yields the nonlinear Maxwell equations
\begin{equation}
    \nabla_\mu \left(\mathcal{L}_{\mathcal{F} }F^{\mu\nu}\right)=0.
    \label{eq:NLEDEquations}
\end{equation}

Our search for black hole configurations begins with a static and circularly symmetric ansatz for the metric, as given in \eqref{eq:AnsatzMetric}, together with a static (purely electric) ansatz for the gauge field:
\begin{align}
 F_{\mu\nu} =  2\, Z(r) \delta^{t}_{[\mu}\delta^{r}_{\nu]},
\label{ElectricAnsatz}
\end{align}
where $-F_{tr} = Z(r)=\partial_rA_t(r)$ is an arbitrary function to be determined dynamically, and corresponds to the electric field of the configuration.

Evaluating the invariant \eqref{eq:InvariantF} yields $\mathcal{F} = -\frac{Z^2}{2}$. In terms of it, the nonlinear Maxwell equations \eqref{eq:NLEDEquations} have only one non-trivial component. For $\nu = t$, we obtain the equation
$$
\frac{d}{dr}(r\,Z \mathcal{L}_F) = 0,
$$
which admits a first-integral solution
\begin{equation}
    Z = \frac{1}{\mathcal{L}_F} \frac{q}{r},
    \label{eq:Zsol1}
\end{equation}
where $q$ is an arbitrary integration constant related to the electric charge, as we will show in Sec.~\ref{sec:Thermo}. Notice that the relation \eqref{eq:Zsol1} is independent of the blackening function and, for the special case of Maxwell's linear theory, $Z$ takes a Coulombian form.
Thus, Eq.~\eqref{eq:Zsol1} establishes an algebraic relation between the electric field $Z(r)$ and the derivative of the structural function $\mathcal{L}_{\mathcal{F}}$ of the electrodynamics. Once $\mathcal{L}(\mathcal{F})$ is specified, this relation determines the electric profile compatible with the static ansatz; conversely, a prescribed electric profile $Z(r)$ uniquely defines the corresponding derivative of the structural function of the electrodynamics.

Inserting the metric ansatz \eqref{eq:AnsatzMetric}  together with Eq.~\eqref{eq:Zsol1} into the gravitational field equations \eqref{eq:EFENLED}, we obtain a system of two \emph{a priori} independent differential equations.
\begin{subequations}\label{eq:effective_sys}
\begin{align} 
    \frac{f}{r}\left[\dfrac{1}{2}f'+\kappa\,q Z+r(\kappa \mathcal{L}+\lambda)\right] =0,
    \label{eq:EFEEquations1}
    \\
     r^2 \left( \dfrac{1}{2} f''+\kappa  \mathcal{L} +\lambda\right) =0.
    \label{eq:EFEEquations3}
\end{align}
\end{subequations}
The degenerate metric branch $f = 0$ of \eqref{eq:EFEEquations1} will be discarded. Moreover, the seemingly independent equations \eqref{eq:EFEEquations1} and \eqref{eq:EFEEquations3} are, in fact, functionally related through derivation with respect to $r$, and involving again  the constraint \eqref{eq:Zsol1}. Consequently, solving either one is sufficient for our integration strategy.

\subsection{The static solutions of exponential electrodynamics} \label{sec:static_sols}
 
We aim to characterize the static solutions arising from the dynamical system \eqref{eq:effective_sys} in the presence of an exponential family of nonlinear electrodynamics. We consider a general structural form of the exponential which may depend on an arbitrary power of the Maxwell invariant
\begin{align} \label{eq:L_exp}
    \mathcal{L}(\mathcal{F}) = \beta \exp\left[ \gamma (-\mathcal{F})^p  \right],
\end{align}
where $\beta$, $\gamma$,  and $p$ are arbitrary real parameters. After differentiating with respect to $\mathcal{F}$ and using Eq.~\eqref{eq:Zsol1}, one finds the relation
\begin{equation}
    r\, Z^{2p-1} = q\, \frac{2^{p-1}}{p \beta \gamma} \, \exp\left[ \gamma Z^{2p}  \right].
\end{equation}
 In this form, relation \eqref{eq:Zsol1} determines the electric field consistent with the static geometry and governed by the exponential electrodynamics introduced in Eq.~\eqref{eq:L_exp}.

Although, in principle, many solutions $Z = Z(r)$ may exist, the only case admitting a closed-form expression in terms of elementary functions occurs for $p = 1/2$, yielding
\begin{align} \label{eq:Zgen_sol}
Z(r)=\dfrac{1}{\gamma} \ln\left( \dfrac{\gamma \beta}{\sqrt{2}q}\,r \right) ,
\end{align}
being this solution physically acceptable when $\gamma \beta/q>0$.


Now, from \eqref{eq:effective_sys}, and using the result for the electric field derived above, we can integrate the metric function as
\begin{align} \label{eq:fsol_NLED}
    f(r) =  \dfrac{ r^2}{l^2} -M - \frac{2\sqrt{2} \, \kappa q r}{\gamma}   \ln\left( \frac{\beta \gamma r}{\sqrt{2} q} \right),
\end{align}
where $M$ is an arbitrary integration constant and the cosmological constant has been fixed in terms of the $AdS_3$ radius as $\lambda=-1/l^2$. Note however, that the dimensionless argument of the logarithm is fixed here by the combination of couplings and charge, which effectively defines a reference length scale $r_0=\sqrt{2}q/(\beta\gamma)$. This scale is not determined by the AdS geometry, in contrast with the solution \eqref{eq:fFunctionByMGT}, and instead reflects the freedom associated with the additional integration constants of this solution.

Notice that the metric function introduced in \eqref{eq:fFunctionByMGT} is a particular case of \eqref{eq:fsol_NLED} for a specific choice of parameters; this solution describes a BTZ black hole deformed by a linear term in $r$, accompanied by a logarithmic contribution. In the parametrization below, our solution reproduces the quantum-NMG geometry exactly by setting
\begin{align}  \label{eq:parameter_map}
    \beta = -\frac{\alpha}{2 e\kappa l}, \qquad 
    \gamma = -\frac{2\sqrt{2} \kappa q}{\alpha},
    \end{align}
where $e$ denotes Euler’s number, and the remaining parameter $\alpha$ is a short-hand for the mass-dependent function induced by the quantum energy--momentum tensor, defined as
\begin{align} \label{eq:alpha}
    \alpha = \frac{2N \ell_p F(M)}{M l^2}.
\end{align}

In this sense, among the broader four-parameter family of exponential electrodynamics defined in \eqref{eq:L_exp}, the analytically tractable solution to this system is supported by the uniparametric structural function
\begin{align} \label{eq:L_exp2}
    \mathcal{L}(\mathcal{F}) = -\frac{  \alpha}{2\kappa l} \,  \exp\left[ -2\,\kappa\frac{q}{\alpha} (-2\mathcal{F})^{1/2} - 1 \right],
\end{align}
and
\begin{align}
    Z(r) =  \dfrac{1}{2\kappa}\dfrac{\alpha}{q}\left[1- \ln\left( \frac{r }{l}\right)  \right],
\end{align}
is the specific electric field solution to this case.

\section{ Dynamical equivalence between quantum-corrected NMG and Einstein-NLED gravity \label{sec:duality}}

In Sec.~\ref{sec:ExactSolutions}, we showed that the logarithmically deformed BTZ black hole from a quantum origin is quite naturally obtained as an exact solution of Einstein gravity coupled to an exponential family of nonlinear electrodynamics. Many examples of this kind are well known in the literature, where the same geometry emerges within distinct theoretical frameworks. Yet, they are often regarded as mere coincidences and left unexplored. In this work, we take a different route: we carefully study the dynamical equivalence between quantum–backreacted NMG and the NLED that supports  this shared solution. This analysis, in turn, reveals a plausible reinterpretation of the parameters appearing in the quantum version of the black hole from a purely classical perspective.

To make this statement precise, we must first scrutinize the common on–shell content of the two theories under the assumption of stationarity and circularity. Rather than comparing particular solutions case by case, we introduce a minimal set of structures on the solution space that allow us to phrase the overlap as a genuine dynamical equivalence problem. In this sense, the question reduces to characterizing the intersection of the on–shell configurations of the two theories under suitable conditions. We now set up these definitions and derive the corresponding consistency relations that will lead to an explicit form of the dynamical equivalence.

Say $S[g^{\mu\nu}, \Psi^{A}]$ is the action defining a gravitational theory for a metric $g_{\mu\nu}$ and backreacting fields $\Psi^{A}$ of any nature and with a collective index $A$. Define $\mathcal{E}=0$ to be schematically the set of all field equations, including the corresponding Einstein field equations and the matter equations. We say that 
\begin{align}
    \Sigma=\left\{  (g_{\mu\nu}, \Psi^{A}) \, \Big | \, \left. \mathcal{E}\right|_{(g_{\mu\nu}, \Psi^{A})}\equiv0 \right\},
\end{align}
is the space of solutions of the theory $S$, whereas by removing gauge redundancies, we can create the space of equivalence classes of physically distinct solutions denominated \emph{moduli space}, namely
\begin{align} 
    \mathcal{M}=\Sigma / \mathcal{G}, \qquad \mathcal{G}=\text{Diff}\times G,
\end{align}
where $\text{Diff}$ is the diffeomorphism group of the gravitational sector and $G$ the full gauge group of matter theory. 

The result of the preceding section shows that the intersection 
\begin{align}\label{eq:intersec}
\mathcal{I}:=\mathcal{M}_{\text{NMG}+\text{Quantum}} \cap \mathcal{M}_{\text{NLED}}\neq 
\emptyset,
\end{align} 
is non-empty, where $\mathcal{M}_{\text{NMG}+\text{Quantum}}$ is the moduli space of NMG supplemented by the quantum-inspired energy-momentum tensor \eqref{eq:q-Tem}, and $ \mathcal{M}_{\text{NLED}}$ the one for GR coupled to a nonlinear electrodynamics.
This intersection includes the non-vacuum extension of the BTZ black hole \eqref{eq:solution}, suggesting at least an interesting dynamical correspondence between the two theories which can be posed as a degeneracy between their moduli spaces.

Following this idea, the first question to address is how to characterize the full intersection \eqref{eq:intersec} of the moduli spaces under suitable assumptions. More precisely, such an on-shell degeneracy can be formulated directly at the level of the field equations:
\begin{subequations} \label{eq:FEqs_2}
\begin{align}
    G_{\mu\nu} + \lambda g_{\mu\nu} &= \kappa\, T^{\text{NLED}}_{\mu\nu}, \\
    G_{\mu\nu} + \lambda g_{\mu\nu} &= \frac{1}{2m^{2}} K_{\mu\nu} +\kappa  N \langle T_{\mu\nu} \rangle.
\end{align}
\end{subequations}

We now choose a static metric ansatz $g_s \in \mathcal{I}$, in the form given in Eq. \eqref{eq:AnsatzMetric}, which is assumed to belong to both moduli spaces. Accordingly, both equations in \eqref{eq:FEqs_2} are to be satisfied when evaluated on $g_s$. This implies that the following consistency relation must hold:
\begin{align} \label{eq:deg_condition}
    \left. \left( \kappa \,T^{\text{NLED}}_{\mu\nu} 
    - \frac{1}{2m^{2}} K_{\mu\nu} 
    - \kappa N \langle T_{\mu\nu} \rangle \right) \right|_{g_s} = 0.
\end{align}

For precision, we state that the static metric is taken as in \eqref{eq:AnsatzMetric}, the field strength tensor is assumed in the electric form \eqref{ElectricAnsatz}, and the electrodynamics is described by a general structural function $\mathcal{L}(\mathcal F)$. Evaluation of the degeneracy (intersection) condition \eqref{eq:deg_condition} yields a system of three fourth-order differential equations, which must be supplemented by the nonlinear Maxwell equations \eqref{eq:NLEDEquations}. A convenient strategy is to solve for higher-order derivatives and substitute them back into the system, thereby reducing its complexity. In fact, the equation governing $f(r)$ can be completely decoupled as
\begin{align}
    r f^{(4)} - 2 f''' = 0,
\end{align}
which is a degenerate Euler-type equation with a general solution
\begin{align} \label{eq:EquivClass_g}
    f(r) = c_0 + c_1 r + c_2 r^2 + c_3 r \ln(r),
\end{align}
where $c_{0,1,2,3}$ are arbitrary integration constants. 

The striking outcome of this analysis is the general form of the static metric \eqref{eq:EquivClass_g} that characterizes the equivalence class of the intersection $\mathcal{I}$. We have shown that the logarithmic generalization of the BTZ black hole is, in fact, the most general static solution shared by the quantum-backreacted NMG theory and a self-gravitating nonlinear electrodynamics. Explicitly, this class is given by
\begin{align}
\mathcal{I} =& \Big\{ g = -f\,dt^{2} + f^{-1}dr^{2} + r^2 d\phi \nonumber\\ \,&\Big|\, f(r) = c_0 + c_1 r + c_2 r^2 + c_3 r \ln(r) \Big\}.
\end{align}

In addition to identifying a common geometric observable (the metric) shared by these two theories, the previous result suggests that, within the classical regime, the NLED model provides an effective theory for the logarithmic deviation from the BTZ solution, as compared to the higher-derivative quantum-induced source in the other scenario. This phenomenological insight is further reinforced by the uniqueness of the solution within the considered ansatz,  and provides a natural bridge for interpreting the parameters of the quantum black hole in terms of its classical continuation, an aspect to be further analyzed in Sec. \ref{sec:classical_lim}.


\section{Singularities and Horizon structure \label{sec:horizons} }

As a natural extension of the BTZ family, we anticipate that the configuration defined by Eqs.~\eqref{eq:Zgen_sol}–\eqref{eq:fsol_NLED}, supported by the exponential electrodynamics \eqref{eq:L_exp}, also represents a genuine black hole solution. From this perspective, an essential element that has remained unexplored in the existing literature concerns the detailed characterization of its causal structure---namely, the identification of event horizons and the classification curvature singularities.

To this aim, let us first consider the general form of the metric family given in Eq.~\eqref{eq:EquivClass_g}; later, we may interpret our results in terms of the parameters of the specific solution \eqref{eq:fsol_NLED} supported by the NLED. We begin by examining the three standard curvature invariants,

\begin{widetext}
\begin{align} \label{eq:invariants}
    \text{Ricci scalar:}&\quad \qquad \,\,\qquad R=-6c_{2} - \frac{2c_{1}+ c_{3}\left(3+2\ln(r)\right) }{r}, \\
    \text{Squared Ricci tensor:}&\quad  \quad \,\,\,\,R_{\mu\nu}R^{\mu\nu}= 12c_{2}^{2} + \frac{8c_{1}c_{2}+c_{2}c_{3}\left( 12+8\ln(r) \right) }{r}
 \\ & \nonumber \qquad\qquad\qquad \,\quad+ \dfrac{1}{2}\frac{3c_{1}^{2}
 + 8c_{1}c_{3} + ( 8c_{3}^{2} +   6c_{1}c_{3})\ln(r) +  3c_{3}^{2}(2+ \ln(r)^{2} )}{r^{2}},
\\
\text{Kretschmann scalar:}&\quad R_{\mu\nu\alpha\beta}R^{\mu\nu\alpha\beta}=12c_{2}^{2} + \frac{  8c_{1}c_{2}+c_{2}c_{3}\left(12+8\ln(r) \right)  }{r}
 \\ & \nonumber \qquad\qquad\qquad \quad + \frac{ 2c_{1}^{2}+ 4c_{1}c_{3}  + c_{3}^{2}\left( 3+2 \ln(r)^{2}\right)  +  ( 4c_{1}c_{3} + 4c_{3}^{2})\ln(r)
  }{r^{2}}.
\end{align}
\end{widetext}
The invariants in Eqs.~\eqref{eq:invariants} exhibit a central divergence at $r = 0$, signaling the presence of a curvature singularity, in close analogy with the non-regular core of BTZ-like geometries once deformations are introduced. The logarithmic contributions do not alter the leading divergent behavior near the origin, but they do modify the detailed subleading structure of the invariants. In the asymptotic region, these logarithmic terms enter multiplied by inverse powers of the radius, and are therefore suppressed at large distances. As a consequence, all curvature invariants remain finite as $r \to + \infty$, and the spacetime preserves an asymptotically AdS character, while the non-regularity at the origin persists as an intrinsic feature of the geometry.

Recent works have employed the holonomy formalism to  further characterize the nature of the BTZ  central singularity; see, for instance, Ref.~\cite{Briceno:2024ddc}. In that framework, the singularity is identified as being of the quasi-regular type, encoded in the non-trivial $AdS_{3}$ holonomy rather than in local curvature divergences.

To investigate the presence of event horizons, we follow the standard procedure of identifying the Killing horizons. The relevant symmetry generator in this case is that of stationarity, which can be written as
\[
    k = \partial_t 
    \qquad \Rightarrow \qquad 
    k_\mu k^{\mu} = -f(r),
\]
so that the horizon is located on the region where the Killing vector becomes null, i.e. where $f(r)=0$. 
For the present configuration, however, this condition cannot be solved analytically due to the logarithmic dependence in the metric function. 
Nevertheless, we can still extract meaningful information about the causal structure by analyzing the qualitative behavior of $f(r)$ and its derivatives, even without an explicit expression for the horizon radius in terms of the model parameters.

Let us begin by examining the existence of critical points of $f$. These follow from the condition $f'(r)=0$, which can be solved analytically in terms of the two real branches of the Lambert $W$ function, whose solution branching is discussed in Ref.~\cite{veberivc2012lambert}, with the precise form determined by the signs and relative magnitudes of $c_{2}$ and $c_{3}$.
  We summarize the three different possibilities in Table~\ref{tab:criticalpoints}, together with the corresponding expressions for the critical points.
    \begin{table*}[t]
\centering
\setlength{\tabcolsep}{12pt}      
\renewcommand{\arraystretch}{1.30} 

\begin{tabular}{lll}
\hline\hline
\emph{Number of critical points} & \emph{Condition} & \emph{Expression} \\
\hline
\\
One 
& $\displaystyle  \frac{c_{2}}{c_{3}} > 0$
& $\displaystyle 
r_{c}
= \frac{c_{3}}{2c_{2}}\,
W_{0}\!\left(
\frac{2c_{2}}{c_{3}}\,e^{-(1 + c_{1}/c_{3})}
\right)$
\\[10pt]

Two 
& $\displaystyle 
-e^{-1}
<
\frac{2c_{2}}{c_{3}}e^{-(1+c_{1}/c_{3})}
<0$
& $\displaystyle 
r_{c_{(-1)}}
= \frac{c_{3}}{2c_{2}}\,
W_{-1}\!\left(
\frac{2c_{2}}{c_{3}}\,e^{-(1 + c_{1}/c_{3})}
\right)$
\\[6pt]
&
&
$\displaystyle 
r_{c_{(0)}}
= \frac{c_{3}}{2c_{2}}\,
W_{0}\!\left(
\frac{2c_{2}}{c_{3}}\,e^{-(1 + c_{1}/c_{3})}
\right)$
\\[10pt]

One (degenerate)
& $\displaystyle 
\frac{2c_{2}}{c_{3}}e^{-(1+c_{1}/c_{3})}
= -e^{-1}$
& $\displaystyle 
r_{c} =- \frac{c_{3}}{2c_{2}}$
\\[6pt]

\hline\hline
\end{tabular}

\caption{
Critical points of the blackening function according to the sign and mutual relations of the four integration constants.  
For arguments greater than zero, there is a single real solution with $W_0>0$.  
In the interval $(-1/e,0)$, the Lambert function admits two real branches, $W_{-1}<W_{0}<0$ (meaning $r_{c_{(-1)}}>r_{c_{(0)} }$), which give rise to two distinct critical points.  
At the endpoint $-1/e$, both branches merge at $W=-1$, yielding a degenerate critical point.
}
\label{tab:criticalpoints}
\end{table*}
To ensure an asymptotically AdS configuration, two minimal requirements must be satisfied.  
First, the outermost critical point must correspond to a minimum of the blackening function.  
This translates into the condition
\[
2\,c_{2}+\frac{c_{3}}{r_{c}}\geq 0 ,
\]
which guaranties that $f$ is eventually positive after the last horizon.  
Second, the leading behavior of the metric function at large radius must be dominated by a positive $\sim r^{2}$ term.  
From the asymptotic expansion,
\[
f(r\to\infty)=c_{2}r^{2}+\mathcal{O}(r),
\]
we see that imposing $c_{2}>0$ is sufficient to recover the standard AdS falloff.  
When both requirements are met, the configuration indeed describes a well-behaved asymptotically AdS black hole.

\begin{figure*}[t]
    \centering

    \begin{minipage}{0.45\textwidth}
        \centering
        \includegraphics[width=\linewidth]{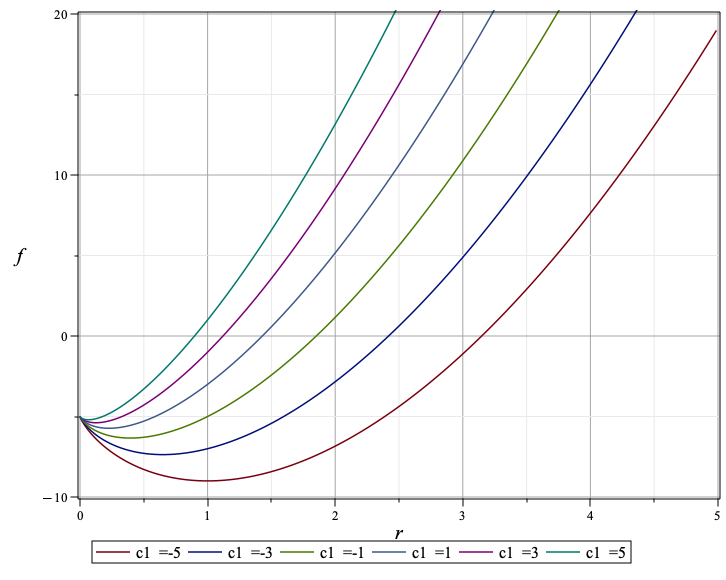}
        \vspace{2mm}
        \textbf{(a)}
    \end{minipage}
    \hfill
    \begin{minipage}{0.45\textwidth}
        \centering
        \includegraphics[width=\linewidth]{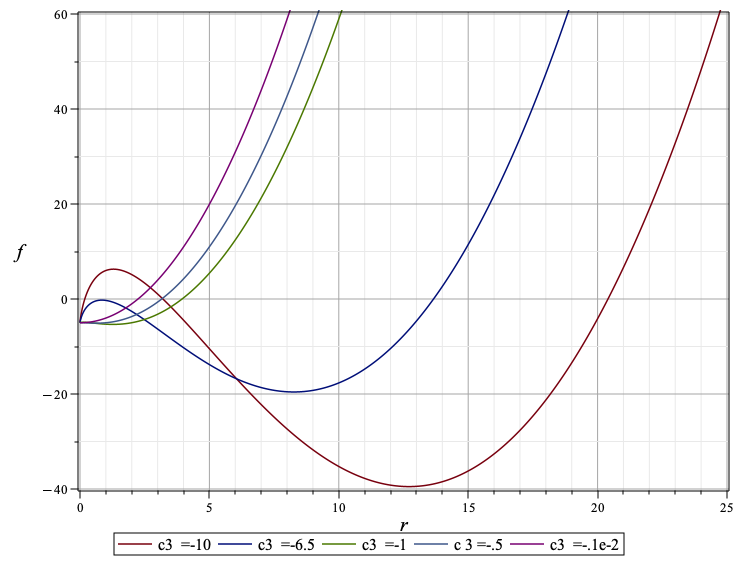}
        \vspace{2mm}
        \textbf{(b)}
    \end{minipage}

    \vspace{6mm}

    \begin{minipage}{0.45\textwidth}
        \centering
        \includegraphics[width=\linewidth]{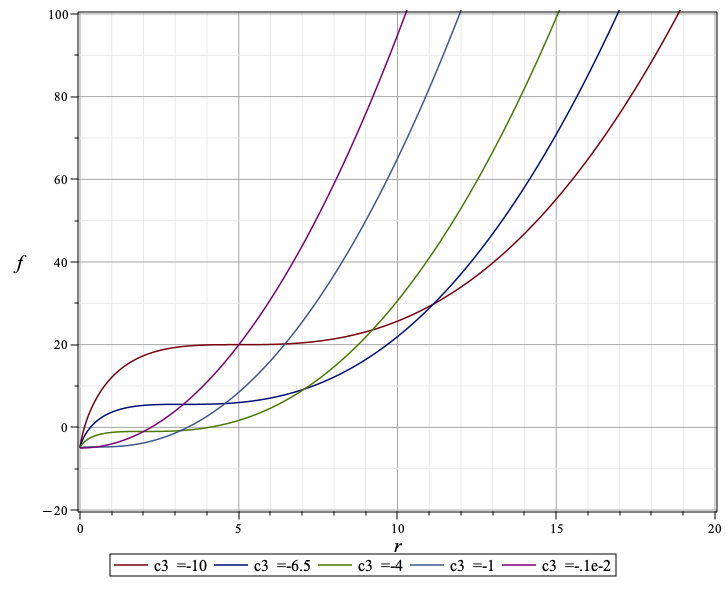}
        \vspace{2mm}
        \textbf{(c)}
    \end{minipage}

    \caption{ Horizon structure in the three possible classes of critical points.  
(a) When a single critical point is present, the solution develops only one horizon, and the sign of the Killing norm reverses once this surface is crossed.  
(b) For configurations admitting two critical points—associated with the two real branches of the Lambert function—the metric can exhibit one, two, or three horizons. In the three-horizon case, the time direction is restored after the first Cauchy horizon, but reverses again upon entering the innermost region.  
(c) In the degenerate case, where the two critical points degenerate into one, the geometry again has a single horizon, with the same reversal of the time direction beyond it.  
In all panels we have fixed $c_{2}=1$ to enforce AdS asymptotics and chosen $c_{0}=-5$, corresponding to a positive mass as discussed in Sec.~\ref{sec:Thermo}.
}
    \label{fig:horizon_cases}
\end{figure*}

In Fig.~\ref{fig:horizon_cases}, we display representative examples of the behavior of the metric function $f(r)$. 
We distinguish three qualitatively different cases according to the structure of its critical points. 
When $f$ admits a single critical point, the solution generically supports a single horizon. 
In the case of two critical points, arising from the two real branches of the Lambert function, the metric function may exhibit one, two, or three horizons, depending on the relative values of the parameters $c_1$ and $c_3$. 
Finally, for the degenerate configuration with a single (double) critical point, the solution again possesses a single horizon.
It would be interesting to explore whether imposing suitable energy conditions on the energy-momentum tensor of the nonlinear electrodynamics further restricts the allowed configurations to the single-horizon case, in analogy with the mechanism recently identified in~\cite{Hale:2025ezt}, where causal NLED theories with finite self-energy are shown to generically eliminate inner Cauchy horizons in charged black holes. See also the discussion in the recent work \cite{Babaei-Aghbolagh:2025tim}.

According to the previous analysis, it is natural to present the horizon structure of the specific semiclassical configuration reported in \cite{Chernicoff:2024dll}. A first inspection of the conditions summarized in Tab.~\ref{tab:criticalpoints} reveals that $c_{2}/c_{3}=1/(\alpha l^{2})>0$ (recalling the definitions in \eqref{eq:alpha}), which places the quantum–corrected geometry in the first case of a single horizon, shown in panel~(a) of Fig.~\ref{fig:horizon_cases}. This behavior reflects the nature of the small parameter $\alpha$, which only induces a mild shift of the horizon radius relative to the BTZ value, without generating inner horizons—contrary to the broader family of charged black hole supported by the nonlinear electrodynamics.

\section{\label{sec:Thermo} Black hole thermodynamics}

\subsection{\label{sec:charges}Charges and the first law}
In order to pave the way toward understanding the dynamical equivalence between both frameworks, a natural question is how the quantum deformation of the geometry leaves its imprint on the thermodynamic quantities. After all, for black holes most of the physically relevant information is encoded in their thermodynamics, ranging from the laws of black hole mechanics to hints about possible holographic dual descriptions. It is therefore worthwhile to establish in detail the thermodynamic properties of the general configuration \eqref{eq:fsol_NLED} and to investigate how the quantum BTZ extensions \cite{Chernicoff:2024dll,Chernicoff:2024lkj} fit within that structure. In what follows we compute the full set of thermodynamic quantities using the Iyer--Wald formalism \cite{Wald1993,IyerWald1994}, which offers a clean and covariant framework to identify the relevant conserved charges.


For convenience, we employ the metric function in the form \eqref{eq:fsol_NLED}, expressed in terms of the electrodynamics couplings and the relevant integration constants. This representation makes the effective parameter space more transparent and already incorporates the conditions required for a well-defined asymptotic behaviour. The Hawking temperature then follows from \eqref{eq:ApWaldForm13} and \eqref{eq:ApWaldForm11}, taking the form
\begin{equation}
    T_{H}
    =\frac{\, r_{h}}{2\pi l^2}
    -\frac{\sqrt{2}\,\kappa q}{2\pi\gamma}
    \left[\ln\!\left(\frac{\beta\gamma r_{h} }{\sqrt{2}\, q}\right)-1\right],
    \label{eq:HawkingTemperature}
\end{equation}
where $r_{h}$ denotes the location of the outermost (event) horizon. The temperature acquires a nontrivial contribution from the logarithmic term, which encodes the dependence on the additional integration constants and reflects the possibility of multiple horizons in the general solution. On the other hand, since the gravitational dynamics is governed by standard Einstein gravity, the entropy follows directly from the Bekenstein area law (see Eq.~\eqref{eq:ApWaldForm13} with $D=3$), yielding
\begin{equation}
    S=\frac{4\pi^{2} r_{h}}{\kappa},
    \label{eq:BekensteinEntropy}
\end{equation}
which coincides with the entropy of the classical BTZ black hole. Moving now to the quantities associated with the gauge sector, the global electric charge is found to be
\begin{equation}
    Q=2\pi q,
    \label{eq:Thermodynamic_charge}
\end{equation}
while its conjugate electrostatic potential reads
\begin{equation} \label{eq:electric_potential}
    \Phi=-\frac{\sqrt{2}\, r_{h}}{\gamma}
    \left[\ln\!\left(\frac{\beta\gamma r_{h} }{\sqrt{2}\, q}\right)-1\right].
\end{equation}

Now, we turn to the global charge generated by the isometry group of the black hole. In this case, the stationarity gives rise to the global mass of the configuration, which after careful consideration of the Iyer--Wald formalism (see Tab.~\ref{tab:ThermoD}) results in
\begin{equation}
    M_T=\frac{\pi}{\kappa}M = M_{\text{BTZ}}.
    \label{eq:Mass}
\end{equation}
This bold result implies that the thermodynamic mass of our configuration strictly coincides with the BTZ mass, meaning that the effect of the backreaction is encoded in the intensive quantities rather than in the global energy. In the framework of extended thermodynamics, one needs to regard the cosmological constant as a pressure variable with an associated conjugate volume, which together supply the work term in the first law. These quantities are
\begin{equation}
    \mathcal{V}=\pi r_h^{2}, 
    \qquad 
    P=\frac{1}{\kappa l^2}.
    \label{eq:Thermodynamicpressureandvolume}
\end{equation}
Under this extended approach, the remaining coupling constants of the NLED sector must also be varied in the first law, each carrying its own conjugate chemical potential. In particular, for the parameters $\beta$ and $\gamma$ one finds
\begin{align}
    \Psi_\beta &= -\frac{2\pi \sqrt{2}\, q\, r_h}{\beta\,\gamma},
    \label{eq:ConjPotentialc} \\
    \Psi_\gamma &= \frac{2\pi \sqrt{2}\, q\, r_h}{\gamma^{2}}
    \left[\ln\!\left(\frac{\beta\gamma r_h}{\sqrt{2}\, q}\right)-1\right].
    \label{eq:ConjPotentialgamma}
\end{align}

Given the relevant thermodynamic quantities, it is straightforward to verify that they satisfy the extended first law,
\begin{equation}
    \delta M
    =T_{H}\,\delta S
    +\Phi\,\delta Q
    +\mathcal{V}\,\delta P
    +\Psi_{\beta}\,\delta \beta
    +\Psi_{\gamma}\,\delta \gamma,
    \label{eq:extendedfirstlaw}
\end{equation}
supplemented by the intrinsic relation among the parameters imposed by the condition $\delta f(r_h)=0$ \footnote{This relation follows by evaluating $f(r_h)=0$, without having the explicit analytic expression of $r_h$ in terms of the remaining parameters. Since $r_h$ is implicitly determined by them, one can at least characterize how their variations are tied together.}.
 
Additionally, from \eqref{eq:fsol_NLED} we can infer the scaling weights of the couplings, namely $\left[\gamma\right]\sim L$ and $\left[\beta\right]\sim L^{-2}$, which already provides enough information to establish a generalized Smarr formula for this family of black holes. One finds that
\begin{equation}
    T_H S - 2P \mathcal{V} + \gamma \Psi_\gamma - 2\beta\, \Psi_\beta = 0,
    \label{eq:Smarrformula}
\end{equation}
which further generalizes the Smarr relation reported in \cite{Liang:2017kht}.

\subsection{\label{sec:stability} Thermodynamic stability and response}

Having established the extended first law and the corresponding Smarr relation, we may now inspect the local thermodynamic stability of the general charged configuration. Within the extended thermodynamics approach---often referred to as black hole chemistry \cite{Kubiznak:2016qmn}---one deals with an enlarged thermodynamic phase space, which allows for a systematic analysis of the corresponding response functions. In this framework, stability is naturally probed through the behavior of quantities that characterize the response of the system to small fluctuations around equilibrium. In what follows, we analyze the specific heat, the electric permittivity, and the generalized susceptibilities associated with the NLED couplings $\beta$ and $\gamma$. This analysis allows us to establish local thermodynamic stability criteria for the black hole.

To this aim, we shall work in the canonical ensemble, where the extensive variables are held fixed. This choice is particularly natural in the present context, since the global electric charge is a conserved quantity, while the additional couplings $\beta$ and $\gamma$ are treated on the same footing within the extended thermodynamic description.

We start by considering the response of the standard thermal and electric sectors. Since the entropy is linear in the horizon radius (Eq.~\eqref{eq:BekensteinEntropy}), it is convenient to regard $r_h$ as the fundamental variable controlling thermal variations at fixed charge. In particular, the specific heat at fixed charges is defined as
\begin{align}
C_{Q}:=T_H\left(\frac{\partial S}{\partial T_H}\right)_{Q,P,\beta,\gamma}
=\,T_H\,\frac{dS/dr_h}{dT_H/dr_h}.
\end{align}
 Evaluating the entropy and the Hawking temperature~\eqref{eq:HawkingTemperature}, one finds the following expression for the heat capacity,
\begin{align}
C_{Q}
=\frac{8\pi^{3}}{\kappa}\,
\frac{T_H}{
\displaystyle
\kappa P-\frac{2\sqrt{2}\,\pi\, Q}{\gamma \,S}
}\,.
\label{eq:CQ}
\end{align}
In the physical region $T_H\geq0$, local thermal stability requires $C_Q>0$, which is equivalently encoded in the condition
\begin{align}\label{eq:CQ>0}
\frac{dT_H}{dr_h}
=\frac{\kappa}{2\pi }P-\frac{\sqrt{2} \,Q}{  \gamma\,S} 
>0.
\end{align}

We next turn to the electric response. A standard probe in this context is the electric permittivity, which in the present setup is most conveniently evaluated at fixed entropy. The adiabatic permittivity,
\begin{align}
\epsilon_S:=\left(\frac{\partial Q}{\partial \Phi}\right)_{S,P,\beta,\gamma},
\end{align}
is particularly simple to compute, since fixing $S$ amounts to fixing the horizon radius. Using the conserved electric charge~\eqref{eq:Thermodynamic_charge} and the electrostatic potential in Eq.~\eqref{eq:electric_potential}, one readily obtains
\begin{align}
\epsilon_S
=\frac{2\sqrt{2}\,\pi^2\,\gamma\,Q}{\kappa\,S}.
\label{eq:epsS}
\end{align}
It then follows that $\epsilon_S$ is positive provided
\begin{align}
\gamma\,Q >0,
\label{eq:EpsilonS>0}
\end{align}
which ensures stability of the gauge sector.

At this stage, the conditions \eqref{eq:CQ>0} and \eqref{eq:EpsilonS>0} already delimit a nontrivial region of the thermodynamic phase space, ensuring local stability under thermal and charge fluctuations. In the enlarged setup, however, the notion of local stability is in general encoded in the characterization of the thermodynamic Hessian with respect to the full set of fluctuating phase space variables. Anyhow, one may adopt a simpler but sufficient criterion: we inspect stability along the additional coupling directions by studying the corresponding one-dimensional response functions (principal minors) associated with $\beta$ and $\gamma$, while keeping the remaining extensive variables fixed.

Concretely, we define the generalized susceptibilities as the variations of the conjugate potentials with respect to their associated couplings, at fixed $(S,Q,P)$ and with the complementary coupling held fixed, namely
\begin{align}
\chi_{\beta}:=\left(\frac{\partial \Psi_{\beta}}{\partial \beta}\right)_{S,Q,P,\gamma},
\qquad
\chi_{\gamma}:=\left(\frac{\partial \Psi_{\gamma}}{\partial \gamma}\right)_{S,Q,P,\beta}.
\end{align}
Following the same strategy as above, these derivatives can be evaluated directly from Eqs.~\eqref{eq:ConjPotentialc}--\eqref{eq:ConjPotentialgamma}, yielding
\begin{align}
\chi_{\beta}
=\frac{\kappa \,Q\,S}{2\sqrt{2}\,\pi^2\beta^{2}\gamma}.
\label{eq:chi_beta}
\end{align}
Thus, under the physical branch requirement $\gamma Q>0$, the susceptibility $\chi_{\beta}$ is manifestly positive provided $\beta\neq 0$, and no additional obstruction arises from fluctuations along the $\beta$ direction.

\begin{figure}[t]
    \centering
    \includegraphics[width=0.85\linewidth]{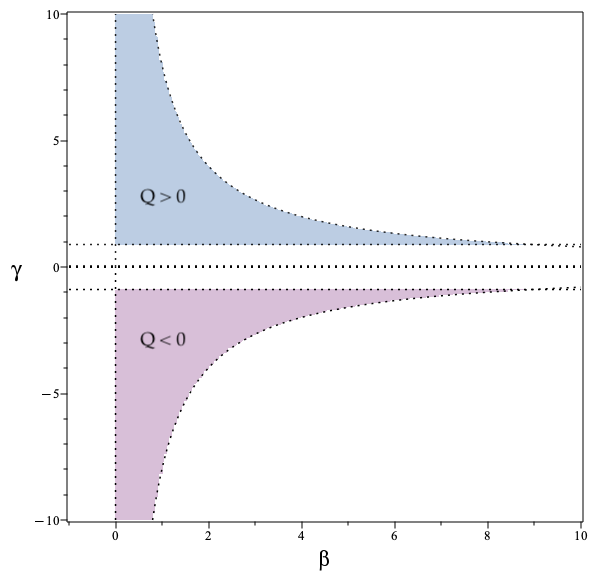}
    \caption{Representative example of the allowed parameter region in the $(\beta,\gamma)$ subspace for which all local thermodynamic stability conditions are simultaneously satisfied. The upper branch corresponds to configurations with positive electric charge ($Q>0$), while the lower branch corresponds to negative charge ($Q<0$). In this plot, the pressure $P$, temperature $T$, and entropy $S$ are held fixed; varying these quantities results only in a rescaling of the admissible region and does not alter its qualitative structure.}
    \label{fig:beta-gamma_region}
\end{figure}

A more involved scenario is found in the $\gamma$ sector, where
\begin{align}
\chi_{\gamma}
=\frac{\kappa \,Q\,S}{2\sqrt{2}\,\gamma^{3}}
\left[\,3-2\ln\!\left(\frac{\kappa\,\beta\,\gamma S}{2\sqrt{2}\,\pi\,Q}\right)\right].
\label{eq:chi_gamma}
\end{align}
Therefore, on the same physical branch $\gamma \,Q>0$, the sign of $\chi_{\gamma}$ is controlled by the logarithmic term. Local stability under $\gamma$--fluctuations is ensured provided
\begin{align}\label{eq:chi_gamma>0}
\,3-2\ln\!\left(\frac{\kappa\,\beta\,\gamma S}{2\sqrt{2}\,\pi\,Q}\right)>0.
\end{align}
Together with Eqs.\eqref{eq:CQ>0} and \eqref{eq:EpsilonS>0}, Eq.~\eqref{eq:chi_gamma>0} provides a compact set of criteria capturing local thermodynamic stability in the enlarged phase space, while keeping the analysis transparent and directly tied to the canonical ensemble. Although the resulting inequalities could be treated analytically, the corresponding expressions are not particularly illuminating. Instead, it is more instructive to visualize the intersection of the stability conditions in a reduced coupling subspace. An explicit example of a fully thermodynamically stable region is displayed in Fig.~\ref{fig:beta-gamma_region} for a representative choice of parameters. The analytical structure of these bounds, as well as their interpretation in the classical-quantum interface, will be revisited in Sec.~\ref{sec:classical_lim}.


\section{Classical limit of the quantum-backreacted BTZ geometry \label{sec:classical_lim}}

We have shown in Sec.~\ref{sec:duality} that the shared solution with the logarithmic correction is not a coincidence, but rather the unique intersection of the dynamical problems of NLED and of NMG supplemented with a semiclassical source, once staticity and circular symmetry are imposed. Along the same lines, one may push the analysis a step further and show that the two parametrizations of the same geometry can be endowed with a deeper physical meaning. In particular, inspecting the mapping between the more general setup of the NLED solution and the parameters entering the quantum-backreacted geometry, given in Eq.~\eqref{eq:parameter_map}, allows one to reinterpret the former as a classical continuation of the latter.

Consider the Lagrangian of the exponential electrodynamics under consideration, rewritten in terms of the quantum-backreacted solution as in Eq.~\eqref{eq:L_exp2}, together with the definition of the parameter $\alpha$ given in Eq.~\eqref{eq:alpha}. One may then perform a series expansion in powers of the relevant ratio $q/\alpha$, keeping $\mathcal F$ fixed. The first terms of the series read
\begin{align}\label{eq:L_series}
\mathcal{L}(\mathcal F)
&= -\dfrac{q}{2\kappa e l}\,\Bigg[
\dfrac{\alpha}{q}
-2\,(-2\mathcal F)^{1/2}
-\dfrac{4\kappa^2 q}{\alpha}\,\mathcal F
+\mathcal{O}\!\left(\frac{q^2}{\alpha^{2}}\right)
\Bigg].
\end{align}
By construction, the parameter $\alpha$ is exponentially suppressed for large values of the mass $M$, and therefore controls the strength of the quantum backreaction on the geometry. It is then natural to explore the behavior of the theory in the two regimes encoded in the expansion \eqref{eq:L_series}, namely when quantum effects dominate and when they are sufficiently suppressed by the black hole scale. For definiteness, let us assume that the ratio $q/(\kappa l)$ is kept fixed throughout the discussion.

In the regime $\alpha/q\gg1$, where quantum effects are dominant, only the first terms in the series \eqref{eq:L_series} are relevant. Effectively, they contribute with a quantum-originated shift to the cosmological constant, proportional to $-\alpha/(2\kappa e l)$, while the main electrodynamic contribution reduces to a square-root theory, $\mathcal{L}\sim\sqrt{-\mathcal F}$. In this regime, any linear Maxwell behavior is largely suppressed. By contrast, in the classical regime $\alpha/q\ll1$, the zeroth-order contribution becomes negligible, and the full exponential structure of the theory is gradually restored. As a result, the expansion \eqref{eq:L_series} contains the linear Maxwell term as part of the subleading contributions, signaling the recovery of standard electrodynamics in the weakly backreacted limit.

Interestingly, even in the classical limit, a remnant of the quantum parameter $\alpha$ persists, together with the ratio $q/\alpha$, both entering as coupling constants of the specific exponential electrodynamics that supports the quantum-backreacted geometry. This naturally raises the question of the physical role played by such parameters once the semiclassical interpretation is adopted. To gain some insight, it is useful to return to the thermodynamic quantities expressed in the parametrization \eqref{eq:parameter_map}, and in particular to the extended-thermodynamics charges $\beta\sim\alpha$ and $\gamma\sim q/\alpha$.

In the classical regime, one has $\beta\to0$, while $\gamma$ survives as an additional global charge characterizing the solution. In contrast, in the quantum-dominated regime the situation is reversed: $\gamma\to0$ whereas $\beta$ remains finite, encoding the strength of the backreaction. This complementary behavior already suggests that $\beta$ and $\gamma$ capture two distinct physical aspects of the same geometry, dominating their thermodynamic contribution in opposite regimes.

This complementary behavior is directly reflected in the thermodynamic sector once evaluated on the parametrization \eqref{eq:parameter_map}. In particular, the Hawking temperature becomes
\begin{equation}
    T_{H}
    =\frac{r_{h}}{2\pi l^2}
    +\frac{\alpha}{4\pi}
    \left[\ln\!\left(\frac{r_{h}}{e\,l}\right)-1\right],
    \label{eq:TH_Qparam}
\end{equation}
which makes explicit how the quantum backreaction modifies the standard BTZ result. In the regime where the system is of a scale comparable to that of $\alpha$, the logarithmic contribution introduces a genuine deviation from the linear BTZ behavior, whereas in the classical limit the usual BTZ temperature is smoothly recovered.

A similar pattern emerges in the response functions. The specific heat, for instance, reads
\begin{equation}
C_{Q}
=\frac{8\pi^{3}}{\kappa}\,
\frac{T_{H}}{\kappa P-\dfrac{2 \pi ^2 \alpha }{\kappa S}},
\label{eq:CQ_Qparam}
\end{equation}
where the effect of the quantum parameter $\alpha$ now appears in the denominator. In the quantum-dominated regime, this contribution opens up the possibility of singular behavior in $C_Q$, signaling a stronger sensitivity of the thermal response to fluctuations. By contrast, in the classical configuration, the denominator reduces to its BTZ counterpart, and the specific heat regains its regular behavior. This neatly mirrors the interpretation of $\alpha$ as a measure of quantum backreaction, whose imprint fades away as the geometry approaches its classical limit.

\section{Concluding remarks \label{sec:conclusions}}

In this work, we have shown that the logarithmically deformed BTZ black hole originally obtained as a quantum--backreacted solution of New Massive Gravity in Ref.~\cite{Chernicoff:2024dll} admits a fully classical realization within Einstein gravity coupled to an exponential model of nonlinear electrodynamics. This correspondence is rooted in a genuine degeneracy of the on--shell dynamics of both theories: once staticity and circular symmetry are imposed, the moduli spaces of quantum--sourced NMG and self--gravitating NLED intersect on a unique class of geometries, characterized by a linear and logarithmic deformation of the BTZ metric. In particular, the solution presented in Eq.~\eqref{eq:fsol_NLED} accommodates two independent parameters beyond the bare BTZ configuration.

An explicit mapping between the parameters of the quantum--sourced black hole and those defining the electrodynamic sector is provided, allowing for a natural identification of nonlinear electrodynamics as an effective classical continuation of the semiclassical description. In this interpretation, the information carried by the quantum sources is encoded in the couplings and conserved charges of the nonlinear electrodynamics. The exponential NLED model uncovered here thus offers a concrete realization in which quantum backreaction can be reinterpreted as a nonlinear self--interaction of classical fields, without the need to invoke higher--derivative corrections in the gravitational sector.

The thermodynamic analysis further reinforces this picture. While the global mass of the black hole remains fixed to its BTZ value, the quantum deformation leaves a nontrivial imprint on intensive quantities and response functions. Within the extended thermodynamic framework, a complementary role between the parameters is uncovered: one of them governs the strength of the quantum backreaction, while the other persists as a genuine classical charge. A careful analysis of local thermodynamic stability further reveals the existence of a physically viable parameter region, as illustrated in Fig.~\ref{fig:beta-gamma_region}.

From the purely classical perspective, we have also examined the horizon structure of the resulting configurations. Although no closed-form expression for the location of the Killing horizons is available, it is nevertheless possible to show that, depending on the relations among the couplings of the theory, as reported in Tab. \ref{tab:criticalpoints}, the gravitational configuration may exhibit one, two, or three horizons, as well as nonphysical naked singularities. On the other hand, the central singularity characteristic of the BTZ geometry persists despite the nonlinear self-interaction of the fields, regardless of the choice of parameters.


Taken together, these results suggest that degenerate black hole geometries admitting both quantum-supported solutions and fully classical realizations can serve as concrete laboratories to explore the classical emergence of a semiclassical spacetime. In this setting, quantum corrections are effectively traded for nonlinearities in the matter sector, rather than for higher-derivative modifications of the gravitational dynamics, as is more commonly assumed. From this perspective, the coexistence of such solutions may be understood as the manifestation of a nontrivial intersection of the on--shell sectors of two distinct theories, hinting at a broader correspondence between curvature corrections of quantum origin and classical nonlinear interactions. Interestingly, related situations have been observed in different contexts, where standard Einstein--matter black hole solutions, such as Kerr--Newman--AdS or Reissner--Nordstr\"om--(A)dS, are reproduced within modified gravity frameworks \cite{Takahashi:2020hso,Minamitsuji:2025cde}, as well as in scenarios involving nonlinear electrodynamics in higher-curvature theories \cite{Singh:2019wpu}. This, in turn, raises the question of whether similar intersections of moduli spaces may occur beyond the static and purely electric sector.

On a complementary front, it would be interesting to investigate whether such classical realizations of semiclassical geometries admit a meaningful extension to black hole chemistry, where variations of the cosmological constant and additional parameters of quantum origin are reinterpreted holographically as changes in the number of degrees of freedom or in the scale of the dual field theory, along the lines discussed in \cite{Karch:2015rpa}.
In this regard, it would be particularly interesting to search for analogous constructions in more general setups, such as nonlinearly charged rotating configurations or solutions accommodating a magnetic charge, like those reported in \cite{Canate:2020btq,Deshpande:2024vbn,Hale:2024zvu,Bravo-Gaete:2025lgs}, completely regular black holes as in \cite{AyonBeatoGarcia1998,AyonBeatoGarcia1999,Bronnikov:2000vy,Fan:2016hvf,Toshmatov:2017zpr,Bueno:2021krl}, or related spacetimes in dimensions higher than three, for instance the analytical solutions reported in \cite{Fernando:2003tz,Hassaine:2007py,Hendi:2012yy,Dymnikova:2015hka,Li:2016nll}. Exploring whether such systems also exhibit a similar overlap between distinct theoretical descriptions would help clarify the extent to which the mechanism identified here persists beyond the present setting. 

\acknowledgments
The authors are grateful to Moises Bravo-Gaete for insightful comments on this work. JAMZ acknowledges partial support through an \emph{Estancias Posdoctorales por M\'exico} grant. ER and JAMZ acknowledge partial support from the \emph{Sistema Nacional de Investigadoras e Investigadores} (SNII) of the Mexican Secretariat for Science, Humanities, and Technological Innovation (SECIHTI). JJSG is partially supported by a \emph{Becas Nacionales para Estudios de Posgrado} grant from SECIHTI. ER also acknowledges encouragement from PRODEP-M\'exico through CA: \'Algebra, Geometr\'ia y Gravitaci\'on.

\appendix

 \section{Details on the Iyer–Wald formalism \label{appx:Iyerwaldformalism}}

An elegant way to extract the thermodynamic properties of a gravitational solution is through the covariant phase space formalism of Iyer and Wald \cite{Wald1993,IyerWald1994}. This framework provides a geometrical and rigorous route to derive conservation laws, charges associated with symmetries, and the laws of black hole thermodynamics in fairly general field theories. 

Given a Lagrangian density $\mathcal{L}(\phi,\nabla\phi,g_{\mu\nu},R_{\mu\nu\rho\sigma})$, where $\phi$ collectively denotes the dynamical fields (including matter sources), we construct the Lagrangian $D$–form $\boldsymbol{L}=\mathcal{L}\,\boldsymbol{\epsilon}$, with $\boldsymbol{\epsilon}$ the natural volume form of the background metric. An arbitrary variation of $\boldsymbol{L}$ yields
\begin{equation}
    \delta \boldsymbol{L}
    =\boldsymbol{E}\cdot \delta \phi
    +d\boldsymbol{\Theta}(\phi,\delta\phi),
    \label{eq:ApWaldForm1}
\end{equation}
where $\boldsymbol{E}$ stands for the equations of motion (a linear $D$–form in $\delta\phi$) and $\boldsymbol{\Theta}$ is the symplectic potential $(D-1)$–form. This identity simply encodes the fact that any variation splits into a term proportional to the field equations plus an exact boundary term. 

In the original formulation, one varies only the fields $g_{\mu\nu}$ and $\phi$ with respect to their functional dependence. In the extended version of the formalism, one also allows variations of the couplings of the theory, including the cosmological constant. In our case of Einstein gravity coupled to NLED we take the Lagrangian density
\begin{equation}
    \mathcal{L}(g_{\mu \nu},A_\mu)
    =\frac{R-2\lambda}{2\kappa}
    -\mathcal{L}_{\text{NLED}}\!\left(\mathcal{F}; \alpha_n  \right),
    \label{eq:ApWaldForm2}
\end{equation}
where $\{\alpha_n\}$ is the set of parameters defining the exponential NLED model. A general variation of $\boldsymbol{L}$ in this extended sense can be written as
\begin{align}
    \delta \boldsymbol{L}
    =\,&\boldsymbol{E}_\phi \cdot \delta \phi
    +d\boldsymbol{\Theta}(\phi,\delta \phi)
    -\frac{\delta \lambda}{\kappa}\boldsymbol{\epsilon}
    +\frac{\partial \mathcal{L}_{\text{NLED}}}{\partial \alpha_m}\delta \alpha_m\,\boldsymbol{\epsilon},
    \label{eq:ApWaldForm3}
\end{align}
where the repeated index $m$ is to be summed over the NLED parameters. 

Under a diffeomorphism generated by a vector field $\boldsymbol{\xi}=\xi^\mu\partial_\mu$, the variation of the fields is given by the Lie derivative, $\delta_\xi \phi=\pounds_\xi \phi$. The corresponding Noether current $(D-1)$–form is
\begin{equation}
    \boldsymbol{J}_\xi
    =\boldsymbol{\Theta}(\phi,\pounds_\xi \phi)
    -i_\xi \boldsymbol{L},
    \label{eq:ApWaldForm4}
\end{equation}
where $i_\xi$ denotes the interior product. On shell this current is exact, $\boldsymbol{J}_\xi=d\boldsymbol{Q}_\xi$, with $\boldsymbol{Q}_\xi$ the Noether potential $(D-2)$–form. For an arbitrary variation $\delta\phi$, one finds
\begin{align}
    d\!\left[\delta \boldsymbol{Q}_\xi-\boldsymbol{\xi}\cdot \boldsymbol{\Theta}(\phi,\delta\phi)\right]
    =\, &\boldsymbol{\omega}(\phi;\delta\phi,\pounds_\xi \phi)
    +\boldsymbol{\xi}\cdot \boldsymbol{E}_\phi\, \delta \phi \nonumber \\
    &-\frac{\delta \lambda}{\kappa}\boldsymbol{\epsilon}
    +\frac{\partial \mathcal{L}_{\text{NLED}}}{\partial \alpha _n}\delta \alpha_n\, \boldsymbol{\epsilon},
    \label{eq:ApWaldForm5}
\end{align}
where $\boldsymbol{\omega}(\phi;\delta\phi,\pounds_\xi \phi)$ is the symplectic current $(D-1)$–form \cite{LeeWald1990}. These ingredients allow one to define the variation of the Hamiltonian charge \cite{IyerWald1994} associated with $\boldsymbol{\xi}$ as
\begin{align}
    \delta H_\xi
    =&\int_{\partial \Sigma}\!\left(\delta \boldsymbol{Q}_\xi-\xi\cdot \boldsymbol{\Theta}(\phi,\delta\phi)\right)
    +\frac{1}{\kappa}\delta\lambda\int_{\Sigma}\!\xi\cdot \boldsymbol{\epsilon} \nonumber \\
    &-\int_{\Sigma}\!\xi\cdot \boldsymbol{\epsilon}\,
    \frac{\partial \mathcal{L}_{\text{NLED}}}{\partial \alpha_n}\delta\alpha_n,
    \label{eq:ApWaldForm6}
\end{align}
where $\Sigma$ is a spacelike hypersurface with boundary $\partial \Sigma$, typically consisting of spatial infinity and/or a section of the horizon. 

We now consider a stationary solution with a Killing horizon generated by $\boldsymbol{\xi}=\partial_t$, together with a linear perturbation $\delta\phi$ solving the linearized equations. In this case, the first integral in \eqref{eq:ApWaldForm6} can be split as
\begin{align}
    \int_{\partial \Sigma}\!\left(\delta \boldsymbol{Q}_\xi -\xi \cdot \boldsymbol{\Theta}(\phi,\delta\phi)\right) 
    =&\int_{S_\infty}\!\left(\delta \boldsymbol{Q}_\xi-\xi\cdot \boldsymbol{\Theta}(\phi,\delta\phi)\right) \nonumber \\
    &-\int_{\mathcal{B}}\!\left(\delta \boldsymbol{Q}_\xi -\xi\cdot \boldsymbol{\Theta}(\phi,\delta\phi)\right),
    \label{eq:ApWaldForm7}
\end{align}
where the first contribution lives at spatial infinity--- and gives the variation of the total energy associated with $\xi$---while the second contribution lives on the horizon $\mathcal{B}$. Since the Killing vector is null on $\mathcal{B}$, one can show that $\left.\left(\xi\cdot \boldsymbol{\Theta}\right)\right|_\mathcal{B}=0$. The integrand then simplifies and \eqref{eq:ApWaldForm6} becomes
\begin{align}
    \delta H_\xi
    =&\,\delta M'
    -\int_{\mathcal{B}}{\delta \boldsymbol{Q}_\xi}
    +\frac{\delta\lambda}{\kappa}\int_{\Sigma}\!\boldsymbol{\xi} \cdot \boldsymbol{\epsilon} \nonumber \\
    &-\int_{\Sigma}\!\boldsymbol{\xi} \cdot \boldsymbol{\epsilon}\,
    \frac{\partial \mathcal{L}_{\text{NLED}}}{\partial \alpha_n}\delta \alpha_n,
    \label{eq:ApWaldForm8}
\end{align}
where $\delta M'$ denotes the contribution from spatial infinity alone.

For the Lagrangian \eqref{eq:ApWaldForm2}, the boundary term associated with the Einstein–Hilbert part is
\begin{equation}
    \Theta^\mu_{\text{EH}}
    =\frac{1}{2\kappa}\left(g^{\mu \alpha}g^{\nu \beta}-g^{\mu \nu}g^{\alpha \beta}\right)\nabla_\nu \delta g_{\alpha \beta},
    \label{eq:ApWaldForm9}
\end{equation}
and using $\delta g_{\alpha \beta}=2\nabla_{(\alpha}\xi_{\beta)}$, the corresponding Noether potential reads
\begin{equation}
    Q^{\mu \nu}_{\text{EH}}
    =-\frac{1}{\kappa}\nabla^{[\mu}\xi ^{\nu]}.
    \label{eq:ApWaldForm10}
\end{equation}

On the horizon, the previous covariant derivative can be expressed in terms of the surface gravity as $\left.\nabla_\alpha \xi_\beta\right|_{\mathcal{B}}=\kappa_s \epsilon_{\alpha \beta}$ \cite{Wald1984}, where $\kappa_s$ is defined by
\begin{equation}
    \kappa_s
    =\left.\sqrt{-\frac{1}{2}\nabla_\alpha \xi_\beta \nabla ^\alpha \xi ^\beta}\right|_{\mathcal{B}},
    \label{eq:ApWaldForm11}
\end{equation}
and $\epsilon_{\alpha \beta}=l_\alpha n_\beta-l_\beta n_\alpha$ is the binormal, with $l_\alpha$ future-directed null and $n_\alpha$ past-directed null. The second term in \eqref{eq:ApWaldForm8} can then be split as $\boldsymbol{Q}_\xi=\boldsymbol{Q}_\xi^{\text{Grav}}+\boldsymbol{Q}_\xi^{\text{NLED}}$, where the purely gravitational piece satisfies
\begin{align}
    \int_{\mathcal{B}}\delta \boldsymbol{Q}_\xi^{\text{Grav}}
    =&\,\frac{1}{2}\delta\int_{\mathcal{B}}{-\frac{1}{\kappa}\nabla^\mu \xi^\nu\epsilon_{\mu \nu}\sqrt{-g}\,d^{D-2}x} \nonumber \\
    =&\,\frac{\kappa_s}{2\pi}\,\delta \left(\frac{2\pi}{\kappa}\Omega_{D-2}r_h^{D-2}\right),
    \label{eq:ApWaldForm12}
\end{align}
where $\Omega_{D-2}$ is the area of the unit $(D-2)$–sphere and $r_h$ is the event-horizon radius. From \eqref{eq:ApWaldForm12} one reads off the Hawking temperature $T_H$ and the Bekenstein entropy $S$ as
\begin{equation}
    T_H=\frac{\kappa_s}{2\pi}, 
    \qquad
    S=\frac{2\pi}{\kappa}\,\Omega_{D-2}r_h^{D-2}.
    \label{eq:ApWaldForm13}
\end{equation}

For the NLED sector, the boundary term associated with \eqref{eq:ApWaldForm2} is
\begin{equation}
    \Theta^\mu_{\text{NLED}}
    =\mathcal{L}_\mathcal{F} F^{\mu \nu}\delta A_\nu,
    \label{eq:ApWaldForm14}
\end{equation}
with $\mathcal{L}_\mathcal{F}=\partial\mathcal{L}_{\text{NLED}}/\partial\mathcal{F}$. To describe the corresponding Noether charge, we consider a combined transformation $\delta_{\xi,\zeta}\phi=\pounds_\xi \phi+\delta_\zeta \phi$ with $\delta _\zeta A_\alpha=\nabla_\alpha \zeta$, where $\zeta$ is a gauge parameter. Using
\begin{equation}
    \pounds_\xi A_\alpha
    =\xi^\beta F_{\alpha \beta}+\nabla_\alpha(\xi ^\beta A_\beta),
\end{equation}
the full variation of the potential becomes
\begin{equation}
    \delta_{\xi,\zeta}A_\alpha
    =\xi^\beta F_{\alpha \beta}+\nabla_\alpha(\xi^\beta A_\beta+\zeta).
    \label{eq:ApWaldForm15}
\end{equation}
The associated NLED Noether charge is then
\begin{equation}
    Q^{\mu \nu}_{\text{NLED}}
    =-\mathcal{L}_\mathcal{F} F^{\mu \nu}\left(\xi ^\beta A_\beta+\zeta\right).
    \label{eq:ApWaldForm16}
\end{equation}

Since the solution is stationary and $A_\mu$ can be chosen so that $\pounds_\xi A=0$ outside the horizon, there exists a gauge parameter $\zeta$ such that $\delta_{\xi,\zeta}A=0$. One can then define the electric conjugate potential at the horizon as
\begin{equation}
\Phi \;:=\; -\,\left.\left(\xi^\alpha A_\alpha\right)\right|_{\mathcal{B}},
\label{eq:ApWaldForm17}
\end{equation}
with $\xi^\alpha A_\alpha \to 0$  at infinity, so that the NLED contribution to \eqref{eq:ApWaldForm8} becomes
\begin{equation}
    \int_{\mathcal{B}}\delta \boldsymbol{Q}^{\text{NLED}}_\xi
    =\Phi\,\delta\!\left(\frac{1}{2}\int_{\mathcal{B}}\epsilon_{\alpha\beta}\mathcal{L_\mathcal{F}}F^{\alpha \beta}\right)
    =\Phi\,\delta Q,
    \label{eq:ApWaldForm18}
\end{equation}
where we identify the electric charge as
\begin{equation}
    Q=\frac{1}{2}\int_{\mathcal{B}}\epsilon_{\alpha \beta}\mathcal{L}_\mathcal{F}F^{\alpha \beta}\sqrt{-g}\,d^{D-2}x
    =\Omega_{D-2}\,q.
    \label{eq:ApWaldForm19}
\end{equation}

Evaluating the Killing generator in \eqref{eq:ApWaldForm8}, and combining \eqref{eq:ApWaldForm13}, \eqref{eq:ApWaldForm17} and \eqref{eq:ApWaldForm19}, we arrive at the extended first law
\begin{equation}
    \delta M
    =T_H \delta S+\Phi\delta Q+\mathcal{V}\delta P+\Psi_{\alpha_n}\,\delta\alpha_n,
    \label{eq:ApWaldForm20}
\end{equation}
where $\mathcal{V}$ is the thermodynamic volume, $P=-\lambda/\kappa$ is the pressure, and $\Psi_{\alpha_n}$ are the conjugate potentials associated with the NLED couplings $\alpha_n$.

For concreteness, the expressions above apply to a spherically symmetric spacetime with a purely electric ansatz in $D$ spacetime dimensions,
\begin{align}
    ds^2=&-f(r)dt^2+\frac{dr^2}{f(r)}+r^2 d\Omega_{D-2}^2, \nonumber \\
    A=&\,A_t(r)\,dt.
    \label{eq:ApWaldForm21}
\end{align}
The resulting thermodynamic quantities that follow from this setup are summarized in Tab.~\ref{tab:ThermoD}.

\begin{table*}[!h]
\centering
\renewcommand{\arraystretch}{1.45}
\begin{tabular}{l l}
\hline\hline
\emph{Thermodynamic Quantity} \hspace{20pt} & \emph{Expression} \\[2mm]
\hline

Mass variation &
$\begin{aligned}
\delta M = -\Omega_{D-2}\lim_{r\to\infty}\Bigg[
&\frac{D-2}{2\kappa}\,r^{D-3}\delta f 
+ A_{t}\,\delta q
+ \frac{\delta\lambda}{\kappa}\frac{1}{D-1}\,r^{D-1} \\
&\hspace{1.2cm}
-\,\delta\alpha_{i}\int\frac{\partial \mathcal{L}_{\mathrm{NLED}}}{\partial\alpha_{i}}\,r^{D-2}\,dr
\Bigg]
\end{aligned}$
\\[6mm]

Hawking temperature &
$\displaystyle 
T_{H}
= \frac{f'(r)}{4\pi}\Big|_{r=r_{h}}
$
\\[4mm]

Electrostatic potential &
$\displaystyle 
\Phi = -A_{t}\Big|_{r=r_{h}}
$
\\[4mm]

Electric charge &
$\displaystyle 
Q = \Omega_{D-2}\,q
$
\\[4mm]

Thermodynamic volume &
$\displaystyle 
\mathcal{V} = \frac{\Omega_{D-2}}{D-1}\,r_{h}^{D-1}
$
\\[4mm]

Pressure &
$\displaystyle 
P = -\frac{\lambda}{\kappa}
$
\\[4mm]

Chemical potentials &
$\displaystyle 
\Psi_{\alpha_{n}}
= \Omega_{D-2}
\int \frac{\partial\mathcal{L}_{\mathrm{NLED}}}{\partial\alpha_{n}}\,
r^{D-2}\,dr\Bigg|_{r=r_{h}}
$
\\[2mm]

\hline\hline
\end{tabular}
\caption{Thermodynamic quantities for a spherically symmetric black hole in $D$ spacetime dimensions, obtained from the extended Iyer–Wald formalism.}
\label{tab:ThermoD}
\end{table*}

The first law has a finite counterpart that relates the thermodynamical quantities. These are also constrained by an underlying scaling structure that manifests even in the extended phase space picture. In particular, once the cosmological constant and the NLED couplings are promoted to thermodynamic variables, the first law acquires a homogeneous character that allows one to infer a corresponding Smarr relation. Following the standard scaling argument \cite{KastorRayTraschen2009}, and assuming that the mass may be regarded as a function of the extensive variables,
$M=M(S,Q,P,\alpha_n)$, one obtains
\begin{align}
    (D-3)M
    =&\,(D-2)T_H S - 2P\,\mathcal{V} \nonumber \\
    &+(D-3)\Phi Q + \sum_n \omega_n\, \Psi_{\alpha_n}\,\alpha_n,
    \label{eq:Apwaldform30}
\end{align}
where the constants $\omega_i$ encode the scaling weights of the NLED couplings, determined by their dimensional assignment,
\begin{equation}
    [\alpha_n]\sim L^{\omega_n}.
    \label{eq:Apwaldform31}
\end{equation}

\section{Plebanski formulation of the supporting NLED \label{appx:Plebanski}} 

An equivalent representation of nonlinear electrodynamics is provided by the Plebanski framework, more commonly referred to as the $\mathcal{H}(\mathcal{P})$ formulation \cite{Plebanski1968}. This approach reformulates the standard $\mathcal{L}(\mathcal{F})$ picture presented in Sec. \ref{sec:ExactSolutions} in a way that is linear in the gauge field $A_{\mu}$, at the price of introducing an auxiliary antisymmetric tensor, the so-called Plebanski tensor $P_{\mu\nu}$. Its relation to the field strength tensor $F_{\mu\nu}$ is dictated by generalized constitutive relations analogous to those of electrodynamics in media,
\begin{equation}
    P_{\mu\nu} = \mathcal{L}_{\mathcal{F}}\,F_{\mu\nu}.
    \label{eq:Ap2Ec1}
\end{equation}

In Plebanski's formulation, the dynamics of the nonlinear electrodynamics is encoded in a structural function $\mathcal{H}(\mathcal{P})$, which is related to the Lagrangian density through a Legendre-type transformation. Explicitly, one has
\begin{equation}
    \mathcal{L}
    =\frac{1}{2}P^{\mu\nu}F_{\mu\nu}-\mathcal{H}
    =2\mathcal{P}\mathcal{H}_{\mathcal{P}}-\mathcal{H},
    \label{eq:Ap2Ec2}
\end{equation}
where $\mathcal{P}=\tfrac{1}{4}P_{\mu\nu}P^{\mu\nu}$ denotes the invariant in Plebanski’s formalism. In what follows, we make use of the constitutive relation \eqref{eq:Ap2Ec2} to construct the dual $\mathcal{H}(\mathcal{P})$ description corresponding to the black hole configuration under consideration. Without loss of generality, let us begin with the relevant theory
\begin{equation}
    \mathcal{L}(\mathcal{F}) = \beta \exp\!\left(\gamma \sqrt{-2\mathcal{F}}\right).
    \label{eq:Ap2Ec3}
\end{equation}
Following Eq. \eqref{eq:Ap2Ec1}, we find the following relation between the invariants
\begin{equation}
    \mathcal{P}=\left(\mathcal{L}_\mathcal{F}\right)^2\mathcal{F},
    \label{eq:Ap2Ec4}
\end{equation}
which, in our case, takes the explicit form
\begin{equation}
    \mathcal{P}=-\frac{1}{2}\beta^2 \gamma^2 \exp\left(2\gamma \sqrt{-2\mathcal{F}}\right).
    \label{eq:Ap2Ec5}
\end{equation}
We may use the reciprocal identity stemming from the constitutive relations
\begin{equation}
    \mathcal{L}_{\mathcal{F}}=\frac{1}{\mathcal{H}_{\mathcal{P}}},
    \label{eq:Ap2Ec6}
\end{equation}
which, for the exponential model under consideration, takes the explicit form
\begin{equation}
   \frac{1}{\mathcal{H}_{\mathcal{P}}}=- \frac{\beta \gamma}{\sqrt{-2\mathcal{F}}}
    \exp\!\left(\gamma \sqrt{-2\mathcal{F}}\right)
    .
    \label{eq:Ap2Ec7}
\end{equation}
Inverting Eq.~\eqref{eq:Ap2Ec5} to obtain $\mathcal{F}(\mathcal{P})$, and combining the result with Eq.~\eqref{eq:Ap2Ec6}, one arrives at
\begin{equation}
    \mathcal{H}_{\mathcal{P}}
    =-\frac{1}{\gamma \sqrt{-2\mathcal{P}}}
    \ln\!\left(\frac{\sqrt{-2\mathcal{P}}}{\beta \gamma}\right).
    \label{eq:Ap2Ec8}
\end{equation}

Finally, after integrating Eq. \eqref{eq:Ap2Ec8} as a differential equation with respect to the invariante $\mathcal{P}$, we end up with
\begin{equation}
    \mathcal{H}(\mathcal{P})= \frac{\sqrt{-2\mathcal{P}}}{ \gamma}\left(\ln\left(\frac{\sqrt{-2\mathcal{P}}}{\beta \gamma}\right)-1\right),
    \label{eq:Ap2Ec9}
\end{equation}
which is the Plebanski's form of the theory \ref{eq:Ap2Ec3}, and can be shown to support the same black hole geometry.
\\ \, \\

\section{Intersection of the moduli spaces: derivation details \label{appx:moduli_detail}}

In this appendix we shed further light on the derivation of the result reported in Eq.~\eqref{eq:EquivClass_g}.  
We begin from the degeneracy condition ---namely the intersection between the two moduli spaces--- expressed in Eq.~\eqref{eq:deg_condition}. For convenience we denote this quantity by the tensor
\begin{equation}
    \Omega_{\mu \nu}=\left.\left(\kappa T^{\text{NLED}}_{\mu \nu}-\frac{1}{2m^2}K_{\mu \nu}-\kappa N \langle T_{\mu \nu}\rangle\right)\right|_{g_{s}},
    \label{eq:Ap3Eq1}
\end{equation}
which satisfies
\begin{align}
\Omega_{\mu \nu}\big|_{\mathcal{I}} \equiv 0 ,
\end{align}
when restricted to the intersection of the moduli spaces $\mathcal{I}$.

Notice that $\Omega_{\mu \nu}$ contains contributions from both sectors of the problem.  
On the one hand it depends on the geometry through the metric function $f$ characterizing the circularly symmetric and static ansatz \eqref{eq:AnsatzMetric}, while on the other, it carries information about the electrodynamics, the higher-curvature terms, and the semiclassical matter sector through the quantities $(Z, \mathcal{L},\mathcal{L}_{\mathcal F})$, the graviton mass $m$, and the remaining coupling constants. Owing to the isometry group considered throughout the manuscript, $\Omega_{\mu\nu}$ possesses only diagonal components.

Taking the trace of Eq.~\eqref{eq:Ap3Eq1} yields
\begin{align}
    g^{\mu \nu}\Omega_{\mu \nu}={}&-\kappa \left(2Z^2 \mathcal{L}_\mathcal{F}+3\mathcal{L}\right)
    -\frac{f''\left(rf''-4f'\right)}{16rm^2}=0 ,
    \label{eq:Ap3Eq2}
\end{align}
which provides a relation between the geometric sector and the electrodynamics.  
From Eq.~\eqref{eq:Ap3Eq2}, together with the algebraic relation \eqref{eq:Zsol1}, the electrodynamics Lagrangian $\mathcal{L}$ can be expressed in terms of derivatives of $f$ and the electric field. Substituting this result back into the components of $\Omega_{\mu\nu}$, one obtains the following reduced system:
\begin{widetext}
\begin{subequations}
\label{eq:AP3eqs}
\begin{align}
        \frac{f\left(8Zqr^2\kappa m^2+24\alpha \kappa Nm^2-6ff^{(4)}r^3-3f'''f'r^3+\left(f''\right)^2r^3-6ff'''r^2-f''f'r^2\right)}{24m^2r^3}=0, 
        \label{eq:Ap3Eq4} \\
        -\frac{8Zqr^2\kappa m^2+24\alpha \kappa Nm^2-3f'''f'r^3+\left(f''\right)^2r^3+6ff'''r^2-f''f'r^2}{24m^2fr^3}=0,  
        \label{eq:Ap3E5} \\
        \frac{8Zqr^2 \kappa m^2+24\alpha \kappa Nm^2-3ff^{(4)}r^3-3f'''f'r^3+\left(f''\right)^2r^3-f''f'r^2}{12m^2r}=0 .
        \label{eq:Ap3E6}
\end{align}
\end{subequations}
\end{widetext}

Although these expressions still contain the electric field, their structure allows for a further simplification. In particular, forming the linear combination $f^2\,\eqref{eq:Ap3E5}+\eqref{eq:Ap3Eq4}$, all dependence on the electromagnetic sector cancels out. This leads to a single differential equation that depends exclusively on the metric function,
\begin{equation}
    -\frac{f^2 \left(f^{(4)}r+2f'''\right)}{4rm^2}=0 .
    \label{eq:Ap3E7}
\end{equation}

Equation \eqref{eq:Ap3E7} admits two solution branches. The trivial branch $f=0$ is physically irrelevant and will be discarded. The remaining branch satisfies
\begin{align} \label{eq:APX_4thO}
rf^{(4)}+2f'''=0 ,
\end{align}
whose general solution is
\begin{equation}
    f(r)=c_0+c_1 r+c_2 r^2+c_3 r \ln r ,
    \label{eq:Ap3E8}
\end{equation}
where the constants $c_i\in\mathbb{R}$ are \textit{a priori} arbitrary. 

We thus arrive at a first conclusion: the moduli space degeneracy condition ($\Omega_{\mu\nu}=0$) restricts the geometry to the four–parameter family of metrics given in Eq.~\eqref{eq:Ap3E8}. This family therefore characterizes the sector of geometries lying at the intersection of the moduli spaces of the semiclassical NMG system and the Einstein–NLED theory under the symmetry assumptions adopted in the present work.

In order to solve the remaining system, we may perform the replacement $f^{(4)}=-2f'''/r$ from Eq.~\eqref{eq:APX_4thO}, which effectively amounts to evaluating the equations on the geometric solution, while avoiding the explicit appearance of the integration constants.
As a consequence, Eqs.~\eqref{eq:AP3eqs} become proportional to
\begin{align}
8 &\kappa m^2 q^2 Z + 24 \kappa \alpha N m^2 + r^3 f'' \nonumber \\
&- 3 r^3 f' f'' - r^2 (f')^2 + 6 r^2 f f'' = 0 \,.
\end{align}
The above equation, together with the algebraic relation in Eq.~\eqref{eq:Zsol1} and Eq.~\eqref{eq:Ap3Eq2}, allows for an explicit determination of the remaining unknowns. In particular, the electric field and the electrodynamics Lagrangian density can be expressed implicitly as functions of $r$ in the form
\begin{subequations} \label{eq:ZL_int}
\begin{align}
Z(r) ={}&
-\dfrac{3\,\alpha  N }{q\,r^2}- \dfrac{1}{8\kappa qm^2 r^2}
\Big[
 (r-1)(f')^{2} \nonumber \\ &
- 3 r f'f''' 
+ 6 f f'' \Big],
\\[6pt]
\mathcal{L}(r) ={}&
\dfrac{ 2\alpha N}{r^3}+\dfrac{1}{16 \kappa m^2 r}\Big[ r( f')^2-4rf'\, f'''+8f\, f'' \Big].
\end{align}
\end{subequations}
This concludes the determination of the electromagnetic sector compatible with the intersection condition. In particular, evaluating the tensor $\Omega_{\mu\nu}$ on the geometry \eqref{eq:Ap3E8} together with the supporting functions \eqref{eq:ZL_int} yields $\Omega_{\mu\nu}=0$, thereby showing that, under the static and circular symmetry assumptions, the gravitational sources produced by the quantum-backreacted NMG system and the Einstein--NLED theory coincide within this class of solutions. In this sense, the two descriptions become dynamically equivalent within the sector under consideration.

\bibliography{DP13450_RESUB1}
\bibliographystyle{apsrev4-2}

\end{document}